\documentclass[superscriptaddress,secnumarabic,nobibnotes,aps,prd,showkeys,showpacs,twocolumn,nofootinbib]{revtex4-1}%

\usepackage[nointlimits]{amsmath}
\usepackage{hyperref}
\usepackage{amsfonts}
\usepackage{amssymb}
\usepackage{amsmath}
\usepackage{amsthm}
\usepackage{latexsym}
\usepackage{graphicx}
\usepackage{color}
\usepackage{layout}
\usepackage[english]{babel}
\usepackage[utf8]{inputenc}

\allowdisplaybreaks

\begin{document}

\title{Gravitational waves in theories with a non-minimal curvature-matter coupling}

\author{Orfeu Bertolami}
\email{orfeu.bertolami@fc.up.pt}
\affiliation{Departamento de Física e Astronomia, Faculdade de Ciências da Universidade do Porto \\ and Centro de Física do Porto \\ Rua do Campo Alegre 687, 4169-007, Porto, Portugal}

\author{Cláudio Gomes}
\email{claudio.gomes@fc.up.pt}
\affiliation{Departamento de Física e Astronomia, Faculdade de Ciências da Universidade do Porto \\ and Centro de Física do Porto \\ Rua do Campo Alegre 687, 4169-007, Porto, Portugal}

\author{Francisco S.N. Lobo}
\email{fslobo@fc.ul.pt}
\affiliation{Instituto  de  Astrofísica  e  Ciências  do  Espaço,  Faculdade  de  Ciências  da  Universidade  de  Lisboa,\\
Edifício  C8,  Campo  Grande,  P-1749-016  Lisboa,  Portugal}

\date{\today}

\begin{abstract}
Gravitational waves in the presence of a non-minimal curvature-matter coupling are analysed, both in the Newman-Penrose and perturbation theory formalisms. Considering a cosmological constant as a source, the non-minimally coupled matter-curvature model reduces to $f(R)$ theories. This is in good agreement with the most recent data.
Furthermore, a dark energy-like fluid is briefly considered, where the propagation equation for the tensor modes differs from the previous scenario, in that the scalar mode equation has an extra term, which can be interpreted as the longitudinal mode being the result of the mixture of two fundamental excitations $\delta R$ and $\delta \rho$.
\end{abstract}

\pacs{04.50.Kd, 04.30.-w, 98.80.Jk}

\maketitle

\section{Introduction}

We have just celebrated the centennial of Einstein's General Relativity (GR), one of the most extraordinary theories ever conceived by the human mind. Despite being derived mainly through theoretical criteria of elegance, aesthetics and simplicity, GR has been extremely successful in accounting for the weak-field experimental regime of gravitation (see e.g. Ref \cite{gravitation, cwill} and references therein). One of its outstanding predictions, namely, gravitational waves (GW) -- ripples in the fabric of spacetime -- had a first indirect observational evidence from the energy loss of the binary pulsar PSR 1913+16 discovered by Hulse and Taylor in 1974 \cite{taylorhulse}, and has recently been directly detected from mergers of black holes binaries by the LIGO collaboration \cite{ligo,ligo2,Abbott:2017vtc,ligo4}. Recently, the observation of gravitational waves from a binary of neutron stars followed by its electromagnetic counterpart has been observed \cite{ligo5,coulter}. This amazing discovery has paved the way for a new era in astronomy, astrophysics and fundamental physics, and opened a new window to test the nature of gravity. GR predicts two massless tensor polarizations travelling at the speed of light, where the amplitude is inversely proportional to the distance from the source. However, extensions of GR predict that additional polarisations may propagate with different velocities, attenuations and effective masses. Some of these observables have been highly constrained: an upper bound on the mass of the graviton \cite{Abbott:2017vtc}, lower and upper bounds on the speed of the gravitational wave \cite{ligons,cornish}. In fact, alternative models equivalent to cosmological scalar fields in scalar-tensor theories of gravity are in difficulties to comply to those bounds \cite{baker,creminelli,sakstein,ezquiaga}.

In order to test gravity with GWs \cite{testGW}, it is important to note that modifications of GR may imply anomalous deviations in the propagation of tensor modes. Detecting new gravitational modes would clearly serve, at a fundamental level, as an {\it experimentum crucis} in discriminating theories since this fact would certainly hint in requiring modifications of GR on large scales. More specifically, in higher order extended theories of gravity \cite{11}, containing scalar invariants other than the Ricci scalar, in addition to a massless spin-2 field, spin-0 and spin-2 massive modes are expected, where the latter may consist of ghost modes. This is indicative that much care should be taken in finding consistent solutions together with weak-field and cosmological constraints, in order to render the solutions physically plausible. However, a number of theories exist where the gravitational wave modes are equivalent to those in GR \cite{12}, and techniques should be devised in order to discriminate models \cite{13}. 

An interesting way to distinguish these theories is the application of the Newman-Penrose formalism according to the Petrov classification \cite{eardleya, eardley,NP, petrov}, in addition to the usual perturbation theory. In $f(R)$ theories \cite{Sotiriou,felice}, these two formalisms seemed to be inconsistent, as the scalar breathing mode that appears in the Newman-Penrose formalism \cite{alves} is not present in the perturbative approach \cite{corda}. Later, it was realised that the usual traceless and transverse conditions for the gauge were incompatible and hence one degree of freedom could not vanish \cite{Jetzer}. Nevertheless, in the Palatini formalism, there are only two tensor modes as in GR \cite{alves}. In fact, this is to be expected as the Palatini-type $f(R)$ models are equivalent to Brans-Dicke theories with the parameter $\omega_{BD}=-3/2$. This particular value corresponds to the vanishing kinetic term of the scalar field, which is thus nondynamical, and therefore no additional scalar degree of freedom should appear. 
Thus, gravitational waves are a powerful tool to test gravitational theories, despite some ambiguities that apparently  exist between different formalisms.

The most useful way to address the problem of gravitational waves is in vacuum. In GR and other alternative theories of gravity where only the gravitational sector is modified, one can consider matter sources and extend the analysis resorting to the Green functions' method. Other approaches have also been discussed such as the Campbell-Morgan formalism \cite{ingraham}, the interaction of gravitational waves with matter \cite{cetoli} or the cyclotron damping and Faraday rotation of gravitational waves in collisionless plasmas \cite{gali, servin}. Another issue that has been explored is the inclusion of a cosmological constant term. The field equations lose their residual gauge freedom \cite{bernabeu} and some implications for the physical metric are encountered \cite{ashtekar}.

Some alternative theories of gravity rely on higher-order curvature terms in the action. This is a relevant modification since it allows for a successful model of inflation, namely the Starobinsky's one \cite{starobinsky}. Furthermore, they are also quite useful at late times where they account for dark matter and dark energy unsolved problems without postulating some exotic particles or fluids yet to be discovered (see Refs. \cite{grob,demg} and references therein).
 
Another proposal of alternative theories of gravity extends the $f(R)$ theories by including a non-minimal coupling (NMC) between curvature and matter \cite{nmc}. These theories can mimic dark matter profiles at galaxies \cite{dmgal,Harko:2010vs} and clusters \cite{dmclusters}, modify the Layzer-Irvine and virial theorem \cite{lieq} and are stable under cosmological perturbations \cite{frazao}. They can also account for the late-time acceleration of the Universe \cite{mimlambda, curraccel} and have some consequences for black hole solutions \cite{bh}. Scalar field inflation is modified for large energy densities though it is still compatible with Planck's data \cite{inflation}. NMC also plays a relevant role during preheating \cite{reheating}, and in the sequestering of the cosmological constant \cite{sequestring}. 
As any model in $f(R)$ theory, there are certain conditions which have to be satisfied in order to ensure that the model is viable and physically meaningful. From the point of view of the energy conditions and of their stability under the Dolgov-Kawasaki criterion, the viability of these models was analysed in Ref. \cite{Bertolami:2009cd}.
Specific wormholes solutions were also presented where normal matter satisfies the energy conditions at the throat, and the higher order curvature derivatives of the NMC are responsible for the null energy condition violation, and consequently for supporting the respective wormhole geometries that satisfy the energy conditions \cite{Garcia:2010xb,MontelongoGarcia:2010xd,Bertolami:2012fz}. 

The nonminimal curvature-matter coupling was also presented in the Palatini formalism \cite{Harko:2010hw}. A maximal extension was explored \cite{Harko:2010mv} by assuming that the gravitational Lagrangian is given by an arbitrary function of the Ricci scalar $R$ and of the matter Lagrangian $\mathcal{L}_m$, i.e., $f(R,\mathcal{L}_m)$ gravity. The geodesic deviation, Raychaudhuri equation, and tidal forces and interesting applications were also analysed in $f(R,\mathcal{L}_m)$ gravity \cite{Harko:2012ve}. Furthermore, extensions of $f(R,\mathcal{L}_m)$ gravity were explored by considering the presence of generalized scalar field and kinetic term dependences \cite{Harko:2012hm} and a NMC between the curvature scalar and the trace of the energy-momentum tensor, the so-called $f(R,T)$ gravity \cite{Harko:2011kv}. The latter was further generalized with the inclusion of a term $R_{\mu\nu}T^{\mu\nu}$ \cite{Haghani:2013oma,Odintsov:2013iba}. We refer the reader to Ref. \cite{Harko:2014gwa} for a recent review on the generalized curvature-matter couplings in modified gravity.

Therefore, much attention has been given to the non-minimal curvature-matter couplings and a study of the associated gravitational waves is a relevant issue. In what follows we shall consider the propagation of gravitational waves in a medium dominated by a cosmological constant and by a dark energy fluid with an equation of state parameter $w=-1$. The first case will reduce to a $f(R)$ scenario, as will be discussed, and the dark energy scenario is obviously of relevance given the late time acceleration of the Universe and the impact of the non-minimal matter-curvature coupling.

This work is organised as follows: In Section \ref{nmc}, the NMC model is presented, as well as its linearisation, and the Newman-Penrose formalism is reviewed. In Section \ref{cc}, we study the case of a background dominated by a cosmological constant both in the perturbation and in the Newman-Penrose formalisms. In Section \ref{darkenergy}, the case of a perfect fluid with equation of state parameter $w=-1$ is analysed. Finally, conclusions are drawn in Section \ref{conclusions}.

\section{The NMC alternative theory}\label{nmc}

\subsection{General formalism}

We start considering the action:
\begin{equation}
S=\int d^4x \sqrt{-g} \left[\frac{1}{2}f_1(R)+f_2(R)\mathcal{L}_m \right] ~,
\end{equation}
where we have set $M_P^2=(8\pi G)^{-1} = 1$ and $c=1$.

Varying the action with respect to the metric, yields the field equations:
\begin{equation}
\label{fieldequations}
\left(F_1+2F_2\mathcal{L}_m\right)R_{\mu\nu} -\frac{1}{2}g_{\mu\nu}f_1-\Delta_{\mu\nu}\left(F_1+2F_2\mathcal{L}_m \right) = f_2 T_{\mu\nu} ~,
\end{equation}
where $F_i\equiv df_i/dR$, $\Delta_{\mu\nu}\equiv \nabla_{\mu}\nabla_{\nu} -g_{\mu\nu}\square$, $T_{\mu\nu}$ is the energy-momentum tensor built from the matter Lagrangian density, $\mathcal{L}_m$, as usual by 
\begin{equation}
T_{\mu\nu}=\frac{-2}{\sqrt{-g}}\frac{\delta (\sqrt{-g}\mathcal{L}_m)}{\delta g^{\mu\nu}}~.
\end{equation}

It is straightforward to retrieve GR by setting $f_1(R)=R$ and $f_2(R)=1$.

Taking the trace of Eq. (\ref{fieldequations}) leads to a useful expression:
\begin{equation}
\label{trace}
\square \left(F_1+2F_2\mathcal{L}_m \right) = \frac{2f_1-\left(F_1+2F_2\mathcal{L}_m \right)R+f_2T}{3} ~.
\end{equation}

From the Bianchi identities, we find that the energy-momentum tensor is not covariantly conserved in general:
\begin{equation}
\nabla^{\mu}T_{\mu\nu} = \frac{F_2}{f_2}\left[g_{\mu\nu}\mathcal{L}_m -T_{\mu\nu}\right]\nabla^{\nu}R ~,
\end{equation}
which, for a perfect fluid with four-velocity $u_{\mu}$, yields an extra force in the geodesic equation:
\begin{equation}
f^{\mu}= \frac{1}{\rho+p}\left[\frac{F_2}{f_2}\left(\mathcal{L}_m -p\right)\nabla_{\nu}R+\nabla_{\nu}p\right]V^{\mu\nu}~,
\end{equation}
where the projection operator $V_{\mu\nu}$ is given by $V_{\mu\nu}=g_{\mu\nu}-u_{\mu}u_{\nu}$.

Analogously to the pure $f(R)$ theories \cite{corda}, the trace equation (\ref{trace}) can be seen as a Klein-Gordon equation, but for two fields $\Phi_1\equiv F_1$ and $\Phi_2\equiv2F_2\mathcal{L}_m$, rather than just one, with an effective potential given by:
\begin{equation}
\frac{dV}{d\Phi_1}\equiv \frac{2f_1-F_1R}{3}, \qquad  \frac{dV}{d\Phi_2}\equiv \frac{-2F_2\mathcal{L}_m R+f_2T}{3}~.
\end{equation}
This identification can be made since the NMC theories are equivalent to a two-scalar field model in the Einstein frame, although one of them is not dynamical as it is mixed with the dynamical one \cite{multiscalar}. Notwithstanding, when considering perturbations, as in the next section, this trace equation leads to a mixture of two fundamental perturbations (associated to the fundamental fields), which in the linearised level are decoupled.

\subsection{Linearised NMC theories}

Perturbing the trace equation around a constant curvature background $R_0$, we obtain:
\begin{equation}
3\square \left(\delta f' +\delta h'\right) = -R_0\left(\delta f' + \delta h'\right) + \delta f +\delta h ~,
\end{equation}
where we have defined the following perturbations:
\begin{eqnarray}
\delta f &&\equiv \left(F_1-2F_2\mathcal{L}_m + F_2 T\right)\delta R ~, \\
\delta f' &&\equiv  \left(F_1'+2F_2'\mathcal{L}_m \right)\delta R ~, \\
\delta h &&\equiv  f_2\delta T ~, \\
\delta h' &&\equiv 2F_2\delta \mathcal{L}_m ~. 
\end{eqnarray}

Linearising the field equations with the metric $g_{\mu\nu}=\eta_{\mu\nu}+h_{\mu\nu}$, where $h_{\mu\nu}\ll 1$:
\begin{eqnarray}
&&\left(F_1'\delta R+2F_2'\mathcal{L}_m \delta R+2F_2\delta L\right)R_{\mu\nu}+\left(F_1+2F_2\mathcal{L}_m \right)\delta R_{\mu\nu} \nonumber  \\
&&-\frac{1}{2}\eta_{\mu\nu}F_1\delta R -\frac{1}{2}h_{\mu\nu}f_1 - \left[\nabla_{\mu}\nabla_{\nu}-h_{\mu\nu} \square \right]\left(F_1+2F_2\mathcal{L}_m \right)  \nonumber \\
&& - \left[\nabla_{\mu}\nabla_{\nu}-\eta_{\mu\nu} \square \right]\left(F_1' \delta R+2F_2'\mathcal{L}_m \delta R + 2F_2\delta \mathcal{L}_m \right)  \nonumber \\
&&=f_2\delta T_{\mu\nu}+F_2T_{\mu\nu}\delta R ~.
\end{eqnarray}

If the background metric is Minkowski, then $R_{\mu\nu}=R=0$ (at lowest order). Furthermore, for constant curvatures (in this case null) we have $\nabla_{\mu}F_i=F_i'\nabla_{\mu}R=0$. Thus: 
\begin{eqnarray}
&&\left(F_1+2F_2\mathcal{L}_m \right)\delta R_{\mu\nu}-\frac{1}{2}\eta_{\mu\nu}F_1\delta R -\frac{1}{2}h_{\mu\nu}f_1   \nonumber  \\
&&- \left[\partial_{\mu}\partial_{\nu}-\eta_{\mu\nu} \square \right]\left(F_1' \delta R+2F_2'\mathcal{L}_m \delta R + 2F_2\delta \mathcal{L}_m \right)   \\
&& + \left[\partial_{\mu}\partial_{\nu}-h_{\mu\nu}\square\right]\left(F_1+2F_2\mathcal{L}_m \right)
    =f_2\delta T_{\mu\nu}+F_2T_{\mu\nu}\delta R\nonumber ~.
\end{eqnarray}

We shall require that $\mathcal{L}_m \to \mathcal{L}_m+\delta\mathcal{L}_m (x^{\mu})$, so that $\nabla_{\mu}\mathcal{L}_m=0$, which is the case of a constant matter Lagrangian density, such as a cosmological constant.
These assumptions further simplify the linearised field equations:
\begin{eqnarray}
&&\left(F_1+2F_2\mathcal{L}_m\right)\delta R_{\mu\nu}-\frac{1}{2}\eta_{\mu\nu}F_1\delta R -\frac{1}{2}h_{\mu\nu}f_1   \\
&&-\left[\partial_{\mu}\partial_{\nu}-\eta_{\mu\nu} \square \right]\left(\delta f' +\delta h'\right) =f_2\delta T_{\mu\nu}+F_2T_{\mu\nu}\delta R \nonumber ~.
\end{eqnarray}

We point out that the far-field, linearised, vacuum field equations for the NMC yield the same result as in $f(R)$ theories. Therefore, we expect to have six polarisation states \cite{alves}.
When considering matter, the polarisation states can be non-trivial, in particular the longitudinal mode, from the trace equation, can be decoupled into two scalar modes depending on the conditions. In the next section we consider the cosmological constant as a source and study the resulting polarisation states.

\subsection{The Newman-Penrose formalism}

In order to properly analyse the polarisation modes of a gravitational theory, one may resort to the Newman-Penrose quantities, which represent the coefficients of the irreducible parts of the Riemann tensor: ten $\Psi$ functions, nine $\Phi$ functions and a single $\tilde{\Lambda}$ function. These structures are related to the independent components of the Weyl tensor, the Ricci tensor and the Ricci scalar, respectively.

We shall consider henceforth the propagating axis of the gravitational wave as oriented in the $+z$ direction. Using the null-complex tetrads \cite{eardley}:
\begin{eqnarray}
&&{\bf k} = \frac{1}{\sqrt{2}}\left({\bf e}_t+{\bf e}_z\right) ~,  \qquad
{\bf l} = \frac{1}{\sqrt{2}}\left({\bf e}_t-{\bf e}_z\right) ~, \\
&&{\bf m} = \frac{1}{\sqrt{2}}\left({\bf e}_x+i{\bf e}_y\right) ~, \quad
{\bf \bar{m}} = \frac{1}{\sqrt{2}}\left({\bf e}_x-i{\bf e}_y\right) ~, 
\end{eqnarray}
which obey $-{\bf k}\cdot {\bf l}={\bf m}\cdot {\bf \bar{m}}=1$ and ${\bf k}\cdot {\bf m}={\bf k}\cdot {\bf \bar{m}}={\bf l}\cdot {\bf m}={\bf l}\cdot {\bf \bar{m}}=0$, respectively. We should bear in mind that any tensor $T$ is written as $T_{abc\dots}=T_{\mu\nu\lambda\dots}a^{\mu}b^{\nu}c^{\lambda\dots}$, where $a,b,c,\dots$ are vectors of the null-complex tetrad basis $({\bf k}, {\bf l}, {\bf m}, {\bf \bar{m}})$, whilst $\mu, \nu, \dots$ run over the spacetime indices.

The Newman-Penrose quantities in the tetrad basis read:
\begin{eqnarray}
\Psi_0 &&\equiv C_{kmkm}= R_{kmkm}\\
\Psi_1 &&\equiv C_{klkm}= R_{klkm}-\frac{R_{km}}{2}\\
\Psi_2 &&\equiv C_{km\bar{m}l} = R_{km\bar{m}l} + \frac{R}{12} \\
\Psi_3 &&\equiv C_{kl\bar{m}l} = R_{kl\bar{m}l} + \frac{R_{l\bar{m}}}{2} \\
\Psi_4 &&\equiv C_{l\bar{m}l\bar{m}} = R_{l\bar{m}l\bar{m}}\\
\Phi_{00} &&\equiv \frac{R_{kk}}{2}\\
\Phi_{11} &&\equiv \frac{R_{kl}+R_{m\bar{m}}}{4}\\
\Phi_{22} &&\equiv \frac{R_{ll}}{2}\\
\Phi_{01} &&\equiv \frac{R_{km}}{2} = \Phi_{10}^* \equiv \left(\frac{R_{k\bar{m}}}{2}\right)^*\\
\Phi_{02} &&\equiv \frac{R_{mm}}{2} = \Phi_{20}^* \equiv \left(\frac{R_{\bar{m}\bar{m}}}{2}\right)^*\\
\Phi_{12} &&\equiv \frac{R_{lm}}{2} = \Phi_{21}^* \equiv \left(\frac{R_{l\bar{m}}}{2}\right)^*\\
\tilde{\Lambda} &&\equiv \frac{R}{24} ~,
\end{eqnarray}
where $C_{\alpha\beta\gamma\delta}$ is the Weyl tensor.

For a metric theory that admits plane null wave solutions, those quantities reduce to only six real independent components in a given null frame: $\{\Psi_2,\Psi_3,\Psi_4,\Phi_{22}\}$, hence one has to consider the ``little group'' $E(2)$, which is a subgroup of the Lorentz group that leaves the wavevector invariant, defining some classes of waves \cite{petrov, eardleya, eardley}. Since both $\Psi_3$ and $\Psi_4$ are complex, each one exhibits two polarisations. This sets under the action of the rotation group yield the helicities $\{0,\pm 1,\pm 2,0\}$. This means that $\Psi_2$ is a longitudinal mode, the real and imaginary parts of $\Psi_3$ account for the mixed vectorial $x-$ and $y-$modes, the $\Psi_4$ denotes the two transverse tensor polarisations ($+,~\times$), and the transverse scalar breathing mode is accounted for $\Phi_{22}$.

In what follows, we shall study the waves arising from a NMC gravity model with a cosmological constant as a source and establish their class.

\section{Cosmological constant as a source}\label{cc}
Let us consider the case of a geometry determined by a cosmological constant :
\begin{equation}
\mathcal{L}_m=-\Lambda \Rightarrow T_{\mu\nu}=-\Lambda g_{\mu\nu} \Rightarrow T=-4\Lambda ~.
\end{equation}
Thus, the field equations read:
\begin{eqnarray}
&&\left(F_1-2F_2\Lambda\right)\delta G_{\mu\nu}-\frac{1}{2}\left(f_1-2f_2\Lambda\right) h_{\mu\nu} \nonumber \\
&&= \left[\partial_{\mu}\partial_{\nu}-\eta_{\mu\nu}\square\right] \left(\delta f' + \delta h'\right) ~,
\end{eqnarray}
where the perturbed Einstein's tensor is defined as $\delta G_{\mu\nu}\equiv \delta R_{\mu\nu}-\frac{1}{2}\eta_{\mu\nu}\delta R$, thus, $R_0=0$, the fluctuations $\delta f'=\left(F_1'-2F_2'\Lambda\right)$ and $\delta h'=0$, since $\delta \Lambda=0$.

The above quantities can be written in terms of the metric perturbations since the perturbed Ricci tensor and scalar curvature read:
\begin{eqnarray}
&&\delta R_{\mu\nu} = \frac{1}{2}\left[-\square h_{\mu\nu}+h_{\nu\; ,\mu\alpha}^{\;\alpha}+h_{\mu\alpha ,\;\nu}^{\;\;\;\;\alpha}-h_{,\mu\nu}\right] ~, \\
&&\delta R = h^{\alpha\beta}{}_{,\alpha\beta} - \square h ~,
\end{eqnarray}
respectively.
 
We choose the following gauge:
\begin{equation}
\label{gauge}
\partial^{\mu}\left[h_{\mu\nu}-\frac{1}{2}\eta_{\mu\nu}h-\eta_{\mu\nu}\left(\frac{\delta f' + \delta h'}{F_1-2F_2\Lambda}\right)\right]=0 ~.
\end{equation}

This is suitable because $F_1-2F_2\Lambda\equiv F_1(R_0)-F_2(R_0)\Lambda$ is computed at a given curvature, in this case $R_0=0$.
In this gauge, we have:
\begin{equation}
\label{gwlambda}
\square \left(h_{\mu\nu}-\frac{1}{2}\eta_{\mu\nu}h-\eta_{\mu\nu}\Omega\right)=\frac{f_1-2f_2\Lambda}{F_1-2F_2\Lambda}h_{\mu\nu} ~,
\end{equation}
with 
\begin{equation}
\Omega\equiv \frac{\delta f'+\delta h'}{F_1-2F_2\Lambda} ~.
\end{equation}

This is interesting as in the adopted gauge, we can have a massive degree of freedom (scalar mode), but we also have a dressed graviton on the RHS that breaks residual gauge invariance $h'_{\mu\nu}=h_{\mu\nu}-\partial_{(\mu}\epsilon_{\nu)}$, as expected \cite{bernabeu}.
It is worth mentioning that the presence of a cosmological constant as an integration constant in the gravity sector \cite{ccint,ccint2,ccint3}, that is $f_1(R)=R-2\Lambda$, leads to a rather different result.

Equation (\ref{gwlambda}) can be cast in a different form:
\begin{eqnarray}
\square \left(h_{\mu\nu}-\frac{1}{2}\eta_{\mu\nu}h\right)=\frac{f_1-2f_2\Lambda}{F_1-2F_2\Lambda}h_{\mu\nu} 
+\frac{\eta_{\mu\nu}}{3}\left(h^{\alpha\beta}{}_{,\alpha\beta} - \square h\right)~.
\end{eqnarray}

However, $\square \Omega = \frac{1}{3}\delta R = \frac{1}{3}\left[-\frac{1}{2}\square h + \square \Omega\right]\Rightarrow \square\Omega=-\frac{1}{4}\square h$. Thus, the previous equation becomes:
\begin{equation}
\label{graviton}
\square\left(h_{\mu\nu}-\frac{1}{4}\eta_{\mu\nu}h\right)=\frac{f_1-2f_2\Lambda}{F_1-2F_2\Lambda}h_{\mu\nu} ~,
\end{equation}
that is, the scalar mode is completely absorbed into a scaling of the term of the trace of the graviton, though it still remains in the gauge choice Eq. (\ref{gauge}). 
This equation has the solution:
\begin{equation}
h_{\mu\nu} = A^+e^{ik_{\alpha}x^{\alpha}}e^+_{\mu\nu}+A^{\times}e^{ik_{\alpha}x^{\alpha}}e^{\times}_{\mu\nu}~,
\end{equation}
where $A^+,~A^{\times}$ are the amplitudes of the ``plus'' and ``cross'' polarisations, and $e^+_{\mu\nu}, ~e^{\times}_{\mu\nu}$ are the usual polarisation tensors, respectively. We further require the following dispersion relations for the tensor modes: $k_{\alpha}k^{\alpha}\equiv\omega^2-k^2=\frac{f_1-2f_2\Lambda}{F_1-2F_2\Lambda}$. This implies that $h=0$, so that the traceless solution naturally occurs when considering a cosmological constant. Consequently, the coefficient of $h_{\mu\nu}$ in Eq. (\ref{graviton}) corresponds to the squared mass of the graviton, which has been recently bounded to $m_g<7.7\times 10^{-23}~ eV/c^2$ \cite{Abbott:2017vtc}. Given the two functions and their first derivatives evaluated at vanishing curvature, and an upper bound on the mass of the graviton, one has that the denominator has to be much larger than the numerator, and that both of them need to have the same sign. For a pure non-minimal matter-curvature coupling ($f_1(R)=R$), and using $\Lambda=4.33\times 10^{-66}~ eV^2$ \cite{planckcc} one has the following restriction: $f_2(0)>-6.8\times 10^{20}$ (in our convention, $c=1$). This is a too feeble constraint on the value of the non-minimal coupling function evaluated at vanishing curvature. Furthermore, in order to avoid tachyonic instabilities, hence $-f_2(0)\Lambda>0\Rightarrow f_2(0)<0$. Thus, a non-minimal coupling to the cosmological constant model is a quite viable model.

The group velocity, $v_g$, of the gravitational wave follows from the dispersion relation, yielding:
\begin{equation}
v_g\equiv\frac{\partial \omega}{\partial k}=\frac{1}{\sqrt{1+\frac{f_1-2f_2\Lambda}{F_1-2F_2\Lambda}\frac{1}{k^2}}}\approx 1-\frac{m_{gw}^2}{2k^2}~,
\end{equation}
where $m_{gw}=\frac{f_1-2f_2(0)\Lambda}{F_1-2F_2(0)\Lambda}\ll 1$, thus this model predicts a group velocity of the gravitational wave slightly smaller that the speed of light, thus avoiding Cerenkov radiation, but really close to it. This is consistent with the most recent constraints $-3\times 10^{-15}<\frac{v_g-c}{c}<7\times 10^{-16}$ \cite{ligons}. Given the dependence of $k$, or equivalently, the energy dependence on the group velocity, the soft-graviton case, where $k\ll 1$, could be problematic. However, for the observed frequencies of gravitational waves $f\sim 250~Hz$, and given the smallness of the graviton's mass, in our model $\omega \sim k$, thus leading to $v_g=1-\frac{m_{gw}}{2k^2}\approx 1- 7.0\times 10^{-23} \to 1^-$, which is compatible with the above bounds.

Furthermore, the so-called "speed" of a gravitational wave, $c_{gw}$, follows from a modified dispersion relation in models with rotation invariance of the form \cite{yunes,cornish}
\begin{equation}
\label{moddisp}
\omega^2=m^2_g+c_{gw}^2k^2+a\frac{k^4}{\Delta}~,
\end{equation}
where $\Delta$ is a high-energy scale cut-off that has been constrained to be large \cite{cutoff}, and $a$ is an operator that depends on each theory. This speed has been constrained to be in the range $0.55<c_{gw}<1.42$ \cite{cornish}. When comparing Eq. (\ref{moddisp}) with the non-minimal coupling dispersion relation, then $c_{gw}=1$, thus being observationally viable.

\subsection{Longitudinal scalar mode}

We shall discuss now the absence of the scalar mode $\Omega$ in the solution of the wave equation. Since the perturbation $\delta h'$ vanishes for a cosmological constant as a source term, the solution for $\Omega$ is straightforwardly achieved. The linearised trace equation gives:
\begin{equation}
\square \Omega = m^2_{\Omega}\Omega ~,
\end{equation}
with 
\begin{equation}
\Omega \equiv \frac{\delta f'}{F_1-2F_2\Lambda} = \frac{F_1'-2F_2'\Lambda}{F_1-2F_2\Lambda}\delta R ~,
\end{equation}
and the mass term 
\begin{equation}
m_{\Omega}^2 \equiv \frac{1}{3}\left[\frac{F_1-2F_2\Lambda}{F_1'-2F_2'\Lambda}\right]  ~.
\end{equation}

In the case of a gravitational wave in a cosmological constant background passing, for instance, through a galaxy, the mass of the fluctuation changes as in $f(R)$ theories \cite{chang}.
Considering a constant matter Lagrangian of the form $\mathcal{L}=-\rho_0$, the underlying physics would be the same and we have only to replace $\Lambda$ by $\rho_0$. Other forms for the matter Lagrangian yield technical problems to handle, however, it does not change our conclusions given the weakness of the coupling of gravity to matter.

\subsection{Newman-Penrose analysis} \label{app:NPformalism}

Now we consider the Newman-Penrose (NP) formalism in order to characterise the full nonlinear NMC model with a cosmological constant background. This case is the same as a $f(R)$ theory with $f(R)\equiv f_1(R)-2 f_2(R)\Lambda$. This means that there may be six polarisation states.

Let us consider, for instance, the following forms for the two functions $f_1(R)=R-\alpha R^{-\beta}$ and $f_2(R)=1+\gamma R^n$. From the trace equation we get:
\begin{eqnarray}
R-4\Lambda =3\square\left(\alpha\beta R^{-\beta-1}-2\gamma n\Lambda R^{n-1}\right)  
	\nonumber \\
+ \alpha(\beta+2)R^{-\beta}+2\Lambda \gamma (2-n)R^n  ~.
\end{eqnarray}

Note that for homogeneous and static scalar curvatures ($\nabla_{\mu}R=0$) and for $n=-\beta=2$, the solution is $R=0$, which is the case of a flat spacetime, for intance, as explored in the previous subsection.

For the special case where $n=-\beta=2$, but not for homogeneous and/or static curvatures, then:
\begin{equation}
\square R = -\frac{R-4\Lambda}{6(\alpha +2\gamma \Lambda)}~,
\end{equation}
whose solution is:
\begin{equation}
R(z,t)=4\Lambda + R_0 e^{ik_{\alpha}x^{\alpha}} ~, 
\end{equation}
with $R_0$ as an integration constant and $\vec{k}=\left(\omega ,0,0,k\right)$ under the condition $k_{\alpha}k^{\alpha}=1/(6\alpha +12\gamma \Lambda)$. The $z$ and $t$ dependences of the Ricci scalar should not be surprising since the computations have been made in the general case without specifying {\it a priori} the spacetime.

Plugging this solution into Eq. (\ref{fieldequations}), we obtain the nonvanishing terms of the Ricci tensor:
\begin{eqnarray}
R_{tt}&&= -\Lambda -\frac{R_0e^{ik_{\alpha}x^{\alpha}}}{2}\left[1-4\left(\alpha+2\gamma\Lambda\right)k^2\right] ~, \\
R_{zz}&&= \Lambda + \frac{R_0e^{ik_{\alpha}x^{\alpha}}}{6}\left[1+12\left(\alpha+2\gamma\Lambda\right)k^2\right] ~, \\
R_{tz}&&= -\left(\alpha+2\gamma\Lambda\right)k \omega R_0e^{ik_{\alpha}x^{\alpha}} ~, \\
R_{xx}&&=R_{yy}=\Lambda+\frac{1}{6}R_0e^{ik_{\alpha}x^{\alpha}} ~.
\end{eqnarray}

It is straightforward to verify that $R=-R_{tt}+R_{xx}+R_{yy}+R_{zz}=4\Lambda+R_0e^{ik_{\alpha}x^{\alpha}}$. In Ref. \cite{alves}, the term $R_{tz}$ was missing, thus leading to some incomplete conclusions.

This is an interesting result, as it yields a Starobinsky-like model with a quadratic non-minimal coupling (the quadratic term yields a GR behaviour with respect to inflation \cite{inflation}).

The nonvanishing NP-Ricci quantities are:
\begin{eqnarray}
\Phi_{00}&&= \frac{\alpha+2\gamma\Lambda}{2}\left(\omega-k\right)^2R_0e^{ik_{\alpha}x^{\alpha}} ~, \\
\Phi_{22}&&= \frac{\alpha+2\gamma\Lambda}{2}\left(\omega+k\right)^2R_0e^{ik_{\alpha}x^{\alpha}} ~, \\
\Phi_{11}&&=\tilde{\Lambda}-\frac{\Lambda}{6} ~.
\end{eqnarray}

This procedure allows us to study the Ricci tensor, or alternatively its traceless version, the Plebański tensor, $S_{\mu\nu}=R_{\mu\nu}-g_{\mu\nu}R/4$, and undertake a classification likewise the Petrov one for the Weyl tensor. As for the Weyl tensor, we must have the full metric in order to compute the Riemann tensor. Expanding the metric in perturbation theory may lead to a different Petrov classification, therefore the full metric is mandatory to fully describe the theory.

For $n=-\beta$, and defining $\phi\equiv R^{-\beta-1}$ the result for the Ricci scalar is as the one obtained in Ref. \cite{alves} with the rescaling
\begin{equation}
\frac{\phi^{-\frac{1}{\beta+1}}-4\Lambda}{\alpha +2\gamma \Lambda}=3\beta \square \phi + (\beta+2)\phi^{\frac{\beta}{1+\beta}} ~.
\end{equation}

In this case, if $\beta \geq 1$, then at late times $R^{-\beta}\gg R, \Lambda$, and the Klein-Gordon equation becomes:
\begin{equation}
\square \phi \approx -\frac{\beta+2}{3\beta}\phi^{\frac{\beta}{1+\beta}} \quad\vee \quad \alpha + 2\gamma \Lambda=0.
\end{equation}

Assuming that $\alpha+2\gamma\Lambda\neq 0$, it is straightforward to obtain the following result:
\begin{equation}
\label{R1}
R(z,t)=\left[i\xi \frac{(z-z_0)-vt}{\sqrt{1-v^2}}+\mathcal{C}^{-1/2}\right]^{-2} ~,
\end{equation}
with 
\begin{equation}
\xi \equiv \frac{1}{2(\beta+1)}\left[\frac{2(\beta+2)(\beta+1)}{3\beta(2\beta+1)}\right]^{1/2}~,
\end{equation}
where $\mathcal{C}$ is an integration constant and $v$ is the wave propagation velocity which arises from the Lorentz transformation. Then the nonvanishing components of the Ricci tensor read:
\begin{eqnarray}
R_{tt}&&=\frac{R}{6\beta}\left[3-2\frac{\beta+2}{1-v^2}\right] ~, \\
R_{xx}&&=R_{yy}=\frac{R}{6\beta}\left[-3+2(\beta+2)\right] ~, \\
R_{zz}&&=\frac{R}{6\beta}\left[-3-2\frac{\beta+2}{1-v^2}v^2\right] ~, \\
R_{tz}&&=\frac{R}{6\beta}\frac{\beta+2}{1-v^2}2v ~.
\end{eqnarray}

Thus, the non-null and independent NP-Ricci quantities are:
\begin{eqnarray}
\Phi_{00}&&=-\frac{R}{12\beta}(\beta+2)\frac{1-v}{1+v} ~, \\
\Phi_{22}&&=-\frac{R}{12\beta}(\beta+2)\frac{1+v}{1-v} ~, \\
\Phi_{11}&&=\frac{R}{12\beta}(\beta+2)=2\frac{\beta+2}{\beta}\tilde{\Lambda} ~.
\end{eqnarray}

The discussion on the NP-Weyl quantities is similar to the previous case.

If $\beta < -2$, under the same conditions as above, then $R^{-\beta}\ll R,  \Lambda$, the Ricci scalar is defined implicitly by:
\begin{eqnarray}
\label{R2}
\frac{(z-z_0)-vt}{\sqrt{1-v^2}}&=\sqrt{6\beta(\alpha+2\gamma\Lambda)}\frac{(\beta+1)}{(\beta+2)}\sqrt{\frac{\beta R^{-1}}{\beta+1}}\times \nonumber\\
&\times ~ _2F_1\left(\frac{1}{2},1+\frac{\beta}{2};2+\frac{\beta}{2};\frac{4\beta\Lambda R^{-1}}{\beta+1}\right) ~,
\end{eqnarray}
where $_2F_1(a,b;c;z)$ is the hypergeometric function \cite{hypergeometric}. This can be evaluated for specific values of the exponent $\beta$. As an example, let us consider the case of $\beta=-3$, then the Ricci scalar reads:
\begin{equation}
R(t,z)=6\Lambda-\frac{\left((z-z_0)-vt\right)^2}{108(\alpha+2\gamma\Lambda)(1-v^2)}
\end{equation}

Thus, the NMC theories with a cosmological constant, or with $\mathcal{L}=-\rho_0=const.$, become a $f(R)$ theory. Furthermore, the polarisation states from the Weyl tensor can be computed from the full metric of the theory. However, the components of the Ricci tensor, the ones computed in this paper, suggest the presence of extra polarisation modes of the gravitational waves of the alternative model. With more data available, the detection of these extra modes could be a direct test of alternative theories of gravity, such as $f(R)$ and the NMC.

\section{Dark-energy-like fluid}\label{darkenergy}

Let us now discuss the case where the perfect fluid matter source has a equation of state parameter of the form $w=-1$, which implies $p=-\rho$:
\begin{equation}
\mathcal{L}=-\rho \approx {\rm const.} \Rightarrow T_{\mu\nu}=-\rho g_{\mu\nu} \Rightarrow T=-4\rho ~.
\end{equation} 

This case resembles the previous one with a cosmological constant, but with a difference, namely that, in general, $\delta \rho \neq 0$, so that the linearised field equations read:
\begin{eqnarray}
\square \left(h_{\mu\nu}-\frac{1}{2}\eta_{\mu\nu}h-\eta_{\mu\nu}\Omega\right)=\nonumber \\
\frac{f_1-2f_2\rho}{F_1-2F_2\rho}h_{\mu\nu} 
+ \frac{2f_2\delta\rho}{F_1-2F_2\rho} \eta_{\mu\nu} ~,
\end{eqnarray}
with 
\begin{equation}
\Omega \equiv \frac{\delta f'+\delta h'}{F_1-2F_2\rho}~.
\end{equation}

Noting that 
\begin{equation}
3\square \Omega = \delta R - \frac{4}{F_1-2F_2\rho}\delta \rho~,
\end{equation}
which means $\square\Omega=-\frac{1}{4}\square h-2\frac{\delta\rho}{F_1-2F_2\rho}$, thus:
\begin{equation}
\square \left(h_{\mu\nu}-\frac{1}{4}\eta_{\mu\nu}h\right)=\frac{f_1-2f_2\rho}{F_1-2F_2\rho}h_{\mu\nu} ~.
\end{equation}

This result is not surprising, as for the cosmological constant case, the trace of the graviton term is reduced by half, but additionally the matter perturbation term disappears due to the fact one required $\rho=\rho_0+\delta\rho$, and the fluctuation is a subleading term.

As far as the longitudinal modes are concerned, and assuming that one can decouple the two modes in the Klein-Gordon equation, which is assuming that matter and curvature perturbations evolve separately at linear order, we have:
\begin{eqnarray}
\square \omega_f = m^2_{\omega_f}\omega_f ~, \\
\square \omega_h = m^2_{\omega_h}\omega_h ~,
\end{eqnarray}
with 
\begin{equation}
\omega_f\equiv \frac{\delta f'}{F_1-2F_2\rho} = \frac{F_1'-2F_2'\rho}{F_1-2F_2\rho}\delta R~,
\end{equation}
and 
\begin{equation}
\omega_h\equiv \frac{\delta h'}{F_1-2F_2\rho} = \frac{-2F_2}{F_1-2F_2\rho}\delta\rho ~.
\end{equation}

Of course, however, this is not the most general case, in which the evolution of matter and curvature perturbations evolve depending on each other, thus being coupled.

The mass terms are defined as 
\begin{eqnarray}
m_{\omega_f}^2 &\equiv & \frac{1}{3}\left[\frac{F_1-2F_2\rho}{F_1'-2F_2'\rho}-R_0\right]~, \\  
m_{\omega_h}^2 &\equiv &\frac{1}{3}\left[\frac{2f_2}{F_2}-R_0\right]~,
\end{eqnarray}
respectively, which in our case $R_0=0$. In order to avoid tachyonic instabilities, one has to require both $\frac{F_1-2F_2\rho}{F_1'-2F_2'\rho} \geq 0$ and $f_2 / F_2 \geq 0$.

Thus, considering a dark energy-like fluid, whose $\mathcal{L}=-\rho\approx const.$,  is quite similar to a cosmological constant in the wave equation for the tensor modes, but with an extra longitudinal mode related to $\delta\rho$.

\section{Conclusions}\label{conclusions}

In this work we have analysed the effects of a non-minimal coupling between matter and curvature on gravitational waves. The perturbation of the trace of the field equations exhibits a behaviour that can be interpreted as the dynamics of the effective scalar field decoupled into two scalar modes: one that arises from perturbations on the Ricci scalar and the other from perturbations on the matter Lagrangian density. In vacuum, the NMC reduces to $f(R)$ and we expect to have the known polarisation modes. 

Considering a background determined by a cosmological constant, we get the same results as in $f(R)$ theories by identifying $f(R)\equiv f_1(R)-2\Lambda f_2(R)$. Nevertheless, through the choice of a suitable gauge, the scalar mode is absorbed into a scaling of the term of $h$ in the wave equation, and the source term given by the graviton is dressed by the functions $f_1(R=0),~f_2(R=0)$ and the cosmological constant. This implies that at linear order the graviton propagates with velocity $v\lesssim c$, and the formalism developed in Refs. \cite{eardleya,eardley} cannot be directly applied \cite{liang}. Instead, one has to compute all the Newmann-Penrose quantities in order to assess the Petrov (for the Weyl tensor) and Plebánski (for the trace-free Ricci tensor) classifications of the theory \cite{WPNP}.

The $f(R)$ and the non-minimal coupling between matter and a cosmological constant differ from GR in the sense that, at linear order, gravitational waves propagate at velocities lower than $c$, and there are extra polarisation modes. In particular, one scalar mode that propagates in the longitudinal way relatively to the gravitational wave has a mass that may be measured.
When considering a gravitational wave passing through the Milky Way's arm where the Solar System lies, assuming that the energy density is roughly constant, $\rho_0$, then the results derived in this paper still apply replacing $\Lambda \to \rho_0$.
If we consider a dark energy-like matter, the propagation equation for the tensor modes is analogous to the one in the cosmological constant scenario, as long as $\rho\approx const.$, but the scalar mode equation has an extra term, which can be interpreted as the longitudinal mode being the result of the mixture of two fundamental excitations arising from $\delta R$ and $\delta \rho$.

Relatively to the detectability of these theories, one may resort to the theory-independent method developed in Ref. \cite{testing}: the antenna angular pattern functions of the detector of gravitational waves measures the linear combination of the polarisation states. Thus, the response from combined Gravitational Wave detectors, such as Advanced-LIGO, Advanced-Virgo or Einstein GW Telescope, could be used to distinguish between models of gravity.

In fact, the most recent bounds on the mass of the graviton and on the velocity of the gravitational waves are in agreement with the non-minimal matter-curvature coupling model.

In what concerns the strong gravity regime, previous work on black hole solutions in the context of curvature-matter nonminimally coupled theories \cite{bh} brings some insight on the issue. Indeed, as discussed in Ref. \cite{bh}, once the Newtonian limit is ensured and the null energy condition is satisfied, it is found that  the Schwarzschild and Reissner–Nordstrom solutions of GR are recovered in the non-minimally coupled curvature-matter theories provided the mass, the charge, and the cosmological constant are suitably ``dressed". This means that existing analyses on gravitational waves generation by black holes collisions will essentially hold with a modification on the relevant parameters so to account for the effects of the non-minimal coupling.

In concluding, the present paper addresses the issue of gravitational waves in NMC theories. Future work requires extending the formalism of cross-correlation analysis so to include the new scalar and vectorial modes. It would then be possible to compute the energy density of the spectrum of a stochastic background of gravitational waves, similar, for instance, to the work of Ref. \cite{detection}.


\section*{Acknowledgments}

The work of CG is supported by Fundação para a Ciência e a Tecnologia (FCT)
under the grant SFRH/BD/102820/2014.
FSNL acknowledges financial support of FCT through an Investigador FCT Research contract, with reference \break IF/00859/2012 and the research grants UID/FIS/04434/2013 and No.~PEst-OE/FIS/UI2751/2014.



\end{document}